\documentclass[twocolumn,prd,amssymb,showpacs,10pt]{revtex4}

\usepackage{graphicx}

\usepackage{bm}

\newcommand{\gsim}{\mbox{\raisebox{-1.ex}{$\stackrel
      {\textstyle>}{\textstyle\sim}$}}}
\newcommand{\lsim}{\mbox{\raisebox{-1.ex}{$\stackrel
      {\textstyle<}{\textstyle \sim}$}}}

\def\tildea{\tilde{r}_A}

\def\luv{l_{\rm uv}}

\def\l4{l_{\rm 4}}
\def\planck5{l_{\rm 5}}

\def\AdS5{\rm AdS_5}

\begin{document}

\title{Holographic Cosmology from the First Law of Thermodynamics 
and the Generalized Uncertainty Principle}
\author{James E. Lidsey}
\affiliation{Astronomy Unit, School of Mathematical Sciences, 
Queen Mary University of London, Mile End Road,
London, E1 4NS, UK}


\begin{abstract}
The cosmological Friedmann equation sourced by the 
trace anomaly of a conformal field theory that is dual to the 
five-dimensional Schwarzschild-AdS geometry can be derived 
from the first law of thermodynamics if the apparent horizon of the 
boundary spacetime acquires a logarithmically-corrected 
Bekenstein-Hawking entropy. It is shown that such a correction 
to the entropy can arise when the generalized uncertainty principle (GUP)
is invoked. The necessary condition for such a thermodynamic derivation 
directly relates the GUP parameter to the conformal 
anomaly. It is consistent with the existence of a gravitational 
cutoff at a scale $\luv \gsim \sqrt{n} \l4$ for a theory containing 
$n$ light species. The absolute minimum in position 
uncertainty can be identified with 
the scale at which gravity becomes effectively five-dimensional. 
\end{abstract}

\vskip 1pc 
\maketitle

A large number of matter fields are predicted to exist in unified 
field theories such as string theory. In the high energy environment  
of the early universe, such fields should behave as a conformal 
field theory (CFT). One-loop quantum corrections break the conformal invariance
of the fields and generate a Weyl (trace) anomaly in the energy-momentum
tensor of the CFT. In general, this is given by 
\begin{equation}
\label{traceanomaly}
g^{\mu\nu} \langle T_{\mu\nu} \rangle = c I_{(4)} - b E_{(4)} ,
\end{equation}
where $I_{(4)} = C_{\mu\nu\lambda\kappa}
C^{\mu\nu\lambda\kappa}$ is the square of the Weyl tensor and
$E_{(4)} = R^2-4R_{\mu\nu} R^{\mu\nu} +R^{\mu\nu\lambda\kappa}
R_{\mu\nu\lambda\kappa}$ is the Gauss-Bonnet invariant. 
(For a review, see, e.g., Ref. \cite{duff}.) The field content of 
the CFT determines the numerical values of the coefficients: 
\begin{eqnarray}
\label{defb}
b= \frac{1}{360 (4\pi )^2} \left( n_0 +11n_{1/2} +62n_1 \right) 
\nonumber \\
c = \frac{1}{120 (4\pi )^2} \left( n_0 +6n_{1/2}+12 n_1 \right)  ,
\end{eqnarray}
where $n_0$, $n_{1/2}$ and $n_1$ are the 
number of scalar, Dirac fermion and vector fields, respectively. 

Conformal field theories 
have a holographic spacetime dual in the large $n$ limit if $b=c$
\cite{hk}. 
For example, the AdS/CFT correspondence 
implies that ${\cal{N}} =4 \hspace{2mm} SU(N)$ super-Yang-Mills 
theory is dual to type IIB string theory on $\AdS5 \times S^5$
\cite{mald,gubser,witten,adsreview}. 
More generally, the properties of the CFT are determined by 
the geometry of the dual spacetime. In particular, the energy-momentum 
tensor of the CFT is determined by means of 
a holographic renormalization scheme \cite{hk,gubser1,hss}. 
For a given solution $G$ to the five-dimensional 
Einstein field equations sourced by a negative cosmological constant, 
$\Lambda_5$, the metric is written in the form  
$ds^2_5 =  \ell^2 z^{-2} [ dz^2 +g_{\mu\nu}
dx^{\mu} dx^{\nu} ]$, where  
\begin{equation}
\label{FGmetric}
g_{\mu\nu} = g_{\mu\nu}^{(0)} + g_{\mu\nu}^{(2)} \frac{z^2}{\ell^2} 
+ g_{\mu\nu}^{(4)} \frac{z^4}{\ell^4} + \ldots \, ,
\end{equation}
and $g_{\mu\nu}^{(i)}= g_{\mu\nu}^{(i)} (x)$ 
solve the gravitational 
field equations. The $z$-coordinate is chosen so that the boundary 
of $G$ is represented by $z=0$. 
It can then be shown that the holographic conformal anomaly
is given by \cite{hk,gubser1,hss}
\begin{equation}
\label{holoanomaly}
\langle g^{(0) \mu\nu} T^{\rm (holo)}_{\mu\nu} \rangle = 
\frac{\ell^3}{128\pi \planck5^3} \left( I_{(4)} - E_{(4)} \right) ,
\end{equation}
where $\ell$ is the curvature radius of $\AdS5$
and $\planck5$ is the five-dimensional 
Planck length. Eq. (\ref{holoanomaly})
corresponds to the standard, four-dimensional, field-theoretic 
result (\ref{traceanomaly}) with $b= c=  \ell^3/(128\pi \planck5^3)$. 

The cosmological consequences of 
such a gauge theory/gravity duality can be investigated 
by parametrizing the line-element (\ref{FGmetric}) 
so that the boundary metric $g^{(0)}_{\mu\nu}$ takes the spatially flat
Friedmann-Robertson-Walker (FRW) form. This metric is made dynamical 
when appropriate mixed boundary conditions are imposed and a 
boundary Einstein action is introduced in the 
holographic renormalization \cite{cm}. The 
effective four-dimensional field equations are then
given by $R_{\mu\nu} - \frac{1}{2}R g_{\mu\nu}^{(0)}  
= 8\pi \l4^2 \langle T^{\rm (holo)}_{\mu\nu} \rangle$, where 
$\l4$ denotes the four-dimensional Planck length. 
Recently, it was shown that 
when the gravity dual is the Schwarzschild-$\AdS5$ geometry, 
the $(00)$-component of the Einstein equations takes the form 
\cite{ast}
\begin{equation}
\label{holographicFriedmann}
H^2 - 16\pi b \l4^2    H^4 = \frac{8\pi \l4^2}{3} \rho  ,
\end{equation}
where $\rho = {\cal{C}} /a^4$ may be interpreted as the energy density of a 
conformally invariant classical fluid, the constant ${\cal{C}}$ 
is determined by the mass of the bulk black hole, $H \equiv \dot{a} /a$
and $a(t)$ denotes the scale factor. 

Eq. (\ref{holographicFriedmann}) can also be derived directly from 
the trace anomaly (\ref{traceanomaly}) for a generic CFT 
(with $b \ne c )$ by integrating the contracted 
Bianchi identity \cite{fhh}. In this case, the parameter ${\cal{C}}$ arises  
as the arbitrary integration constant.

The purpose of the present paper is to show that 
the holographic/conformal-anomaly Friedmann equation 
(\ref{holographicFriedmann}) admits an alternative derivation in terms 
of spacetime thermodynamics and the generalized uncertainty principle 
of quantum gravity. It has long been appreciated that  
a deep connection exists between gravitation, 
quantum theory and statistical physics 
\cite{bek,bch,hawk,israel,bek1,hawktemp}. One of the central themes 
underlying this connection is that a spacetime horizon should 
be associated with an entropy that is directly proportional to 
the horizon area \cite{bek,hawktemp}. 
When such a proportionality exists 
for a local Rindler causal horizon, the Einstein field equations 
can be derived from the fundamental Clausius relation \cite{jac}.  
More specifically, the standard Friedmann equations  
for a spatially isotropic universe 
follow directly from the first law of thermodynamics  
if the entropy and area of the apparent horizon 
satisfy the Bekenstein-Hawking formula, $S= A/(4\l4^2 )$ \cite{caikim}. 

Here we consider the effect of the generalized uncertainty principle 
(GUP) on the apparent horizon entropy. 
The GUP is formulated as the condition 
\begin{equation}
\label{GUP}
\Delta x \Delta p \gsim 1 + \alpha^2 \l4^2 
( \Delta p )^2  ,
\end{equation}
where $\alpha$ is a (model-dependent) dimensionless constant. 
Such a modification to the standard Heisenberg relation  
has been derived in a number of different approaches to quantum gravity, 
including non-commutative quantum mechanics \cite{noncomm} and  
string theory \cite{stringuncert}. 
It also arises from {\em Gedanken} experiments that are 
independent of the underlying theory \cite{maggiore}. 
(For a review, see, e.g., Ref. \cite{garay}.)
We find that 
the GUP induces a logarithmic correction to the entropy that has 
precisely the form required for a thermodynamic 
derivation of the Friedmann equation 
(\ref{holographicFriedmann}). 

To proceed, we adapt a line of reasoning 
developed within the context of black hole spacetimes \cite{mv}.
The spatially flat FRW line element can be written in the 
form $ds^2 =h_{ab}dx^adx^b +\tilde{r}^2 d \Omega^2_2$, where 
the two-metric $h_{ab} = 
{\rm diag} (-1, a^2)$ and $\tilde{r} \equiv ra(t)$. 
The apparent horizon of an observer at $r =0$ is 
the constant-time hypersurface where orthogonal ingoing, future-directed light
rays have zero expansion. It corresponds to a sphere
of radius $\tildea =1/H$ and area 
$A= 4 \pi \tildea^2$, where $\tildea$ is defined by the condition 
$h^{ab} \tilde{r}_{,a} \tilde{r}_{,b} =0$.

In the following, we focus on the regime of cosmic dynamics 
where the universe undergoes a phase of quasi-exponential 
expansion, such that $\dot{\tilde{r}}_A = -\dot{H}/H^2 \ll 1$. 
This implies that over the incremental 
time intervals considered, the 
apparent horizon radius can be regarded as having a fixed value. 
(Such inflationary expansion can be  
realised by introducing an effective four-dimensional cosmological  
constant that is generated by a slowly rolling, self-interacting scalar field. 
We do not exhibit such a term in the Friedmann equations for notational 
simplicity.)

Suppose the apparent horizon absorbs (or emits) a massless quantum 
particle of energy, $\Delta E$. This is determined by the corresponding
uncertainty in the particle's momentum, $\Delta E \simeq \Delta p$ 
\cite{park}.   
The effective mass-energy within the horizon will then change by an amount 
$d{\cal{M}} \simeq \Delta p$. The total mass-energy within the 
apparent horizon is given by ${\cal{M}} = 4\pi \tilde{r}_A^3 \rho /3$ 
and this can be expressed in the 
form ${\cal{M}} = \tildea / (2\l4^2 )$ after substitution of 
the Friedmann equation, $H^2=8\pi \l4^2 \rho /3$. 
(More generally, the horizon mass is the Misner-Sharp  
mass ${\cal{M}} \equiv \tilde{r} [1-h^{ab} \tilde{r}_{,a}
\tilde{r}_{,b} ]/(2\l4^2)$ evaluated at the radius $\tilde{r} =\tildea$
\cite{misnersharp}.)  
The corresponding change in the horizon area as a result of the 
absorption is therefore  
\begin{equation}
\label{changearea}
\Delta A \simeq 16 \pi \l4^2 \tildea \Delta p  .
\end{equation}

The particle will have a Compton wavelength and associated  
uncertainty in position, $\Delta x$. A natural length scale 
for this uncertainty 
is the inverse of the surface gravity at the apparent horizon. 
This is given by $|\kappa |^{-1} = \tildea [1- \dot{\tilde{r}}_A
/2]^{-1} \simeq \tildea$.  
Hence, the position uncertainty can be estimated as  
\begin{equation}
\label{compton}
\Delta x \simeq  \tildea  .
\end{equation}

In this case, it follows that the change in horizon area is  
$\Delta A \simeq (16 \pi \l4^2 ) \Delta x \Delta p$. 
The standard form of the Heisenberg uncertainty principle, 
$\Delta x \Delta p \gsim 1$, would then impose a lower bound 
of $\Delta A \gsim 16 \pi \l4^2 $. 
However, it is straightforward to show that the GUP results in a lower bound 
on the momentum uncertainty (for a given value of $\Delta x$): 
\begin{equation}
\label{deltapmin}
\Delta p \gsim \frac{\Delta x}{2\alpha^2 \l4^2} \left[ 1- 
\sqrt{1- \frac{4\alpha^2\l4^2}{(\Delta x )^2} } \right]   .
\end{equation}
We then deduce that 
\begin{equation}
\label{Delta}
\Delta A \gsim \frac{2 A}{\alpha^2 } \left[ 1 - \sqrt{ 1- 
\frac{16\pi \alpha^2 \l4^2}{A}} \right]   
\end{equation}
after noting that the area of the apparent horizon is 
$A \simeq 4 \pi (\Delta x )^2$. 

If we further assume that the energy scales of interest are sufficiently
small, $\l4^2 /A \simeq  H^2\l4^2  \ll 1$, 
to allow for a consistent Taylor 
expansion of the square root, 
the bound (\ref{Delta}) approximates at leading order to  
\begin{equation}
\label{expansion}
\Delta A \gsim 16 \pi \l4^2
\left[ 1+ \frac{4 \pi \alpha^2 \l4^2}{A} + \ldots 
\right]  .
\end{equation}
This enables us to express the minimum change in the area of 
the apparent horizon as
\begin{equation}
\label{minarea}
(\Delta A )_{\rm min} \simeq 
\epsilon \l4^2 \left[ 1+ \frac{4 \pi \alpha^2 \l4^2}{A} + \ldots 
\right]  ,
\end{equation}
where $\epsilon \simeq {\cal{O}} (16\pi )$ quantifies any further 
uncertainties that may arise \cite{mv}. 

The absorption (or emission) of the particle 
by the horizon results in an increase in the entropy, $\Delta S$. 
Information theory implies that the 
minimal increase should be one `bit of information',  
$(\Delta S)_{\rm min} \simeq b$, where $b \in \Re^+$ and is independent 
of the area. (For a review, see, e.g., Ref. \cite{quantumreview}.)
It follows, therefore, that 
\begin{equation}
\label{diffeqn}
\frac{(\Delta S)_{\rm min}}{(\Delta A)_{\rm min}} 
\simeq \frac{dS}{dA} \simeq \frac{b}{\epsilon \l4^2} 
\left[ 1 - \frac{4 \pi \alpha^2 \l4^2}{A} + \ldots \right]
\end{equation}
and integration of Eq. (\ref{diffeqn}) then yields 
\begin{equation}
\label{entropyarea}
S = \frac{A}{4\l4^2} -  \pi \alpha^2  \ln \left( 
\frac{A}{4\l4^2} \right) + \ldots \, ,
\end{equation}
where we have normalized
$b/ \epsilon = 1/4$ to reproduce the Bekenstein-Hawking 
formula in the limit $\alpha^2 \rightarrow 0$ \cite{mv}. 
In the context of the 
present discussion, this is equivalent to requiring that the 
standard, classical Friedmann equation is recovered in the low-energy 
limit. 

To summarize thus far, the effect of 
the generalized uncertainty principle on the entropy of the  
apparent horizon is to induce  
a leading-order logarithmic correction to the Bekenstein-Hawking 
formula. 

We now apply the first law of thermodynamics, $dE=-TdS$, 
to the apparent horizon. For 
a universe sourced by a perfect fluid with 
energy density, $\rho$, and 
pressure, $P$, the amount of energy, $dE$, crossing the apparent horizon in 
an infinitesimal time interval, $dt$, is evaluated by integrating 
the energy-momentum flux through the horizon and contracting 
with the horizon generator, $k^a=(1,-Hr)$. It can then be shown that 
\begin{equation}
\label{energyflow}
dE =-A T_{ab}k^a k^b dt  = \frac{4\pi}{3H^3} d\rho  , 
\end{equation} 
where it has been assumed implicitly  
that the fluid satisfies the conservation  
equation, $\dot{\rho}=-3H(\rho + P )$.  

If the apparent horizon has an entropy, $S$, and associated Hawking 
temperature, $T = H/ (2\pi )$ \cite{cch}, 
it follows that the first law 
of thermodynamics can be expressed in the form
\begin{equation}
\label{firstlaw}
dS = - \frac{8\pi^2}{3} \frac{d\rho}{H^4}   .
\end{equation}
After substitution of the logarithmically-corrected 
Bekenstein-Hawking entropy (\ref{entropyarea}), 
where $A=4\pi /H^2$, Eq. (\ref{firstlaw}) may be integrated 
to yield \cite{cch1,lidsey}
\begin{equation}
\label{thermoFriedmann}
H^2 - \frac{\alpha^2 \l4^2}{2} H^4 = \frac{8\pi \l4^2}{3} \rho  .
\end{equation}

The effect of the logarithmic correction to the 
entropy is to modify the Friedmann equation from its standard, 
relativistic form. A direct correspondence between the  
holographic/trace-anomaly Friedmann equation 
(\ref{holographicFriedmann}) and the thermodynamic Friedmann equation
(\ref{thermoFriedmann}) is established 
if the GUP parameter, $\alpha$, 
and the anomaly coefficient, $b$, are related such that
\begin{equation}
\label{triality}
\alpha^2 =  32 \pi b = \frac{1}{4} \frac{\ell^3}{\planck5^3}   .
\end{equation} 
The first equality in the correspondence (\ref{triality}) 
holds for a generic CFT, irrespective of 
any holographic considerations, whereas the second arises when 
there exists a spacetime dual to the CFT (i.e., when $b=c$). In a sense, 
Eq. (\ref{triality}) may be regarded as a triality between 
a thermodynamic quantity, $\alpha$, a field-theoretic parameter, $b$, and 
the higher-dimensional gravitational coupling, $\planck5$. 

A question that naturally arises is whether such a correspondence is 
more than an intriguing mathematical analogy. 
It is known that logarithmic 
corrections to the entropy-area law generically arise 
for black hole spacetimes when quantum 
effects are taken into account. Indeed, 
one-loop effects near the event horizon 
of a Schwarzschild black hole lead to a similar proportionality 
between the logarithmic coefficient and the conformal anomaly 
\cite{furs}. This 
suggests that in both the black hole and cosmological environments,  
the conformal anomaly - 
which is geometric in nature and quantum-mechanical in origin - 
may be interpreted in terms of a thermodynamic quantity. 

We may gain further insight by noting that 
an immediate consequence of the GUP   
is that the second term in (\ref{GUP}) results in  
an absolute minimum in the uncertainty in position for any level of 
momentum uncertainty: 
\begin{equation}
\label{minimum}
(\Delta x )_{\rm min} \gsim 2 |\alpha | \l4  .
\end{equation}
Moreover, a generic feature of a theory containing $n$ massless 
fields coupled to gravity is 
the existence of a fundamental length scale, $\luv$, below which low-energy 
perturbation theory is expected to break down. 
Since the GUP originates from quantum gravity considerations,
it is natural to associate the minimum length 
scale with this ultraviolet (UV) cutoff. 
A conservative estimate is that the cutoff occurs at, or on a scale 
slightly above, $(\Delta x)_{\rm min}$, i.e.,  
$\luv \gsim (\Delta x )_{\rm min} \gsim 2|\alpha | \l4$. 
In this case, Eq. (\ref{triality}) implies that 
\begin{equation}
\label{bcutoff}
b \lsim \frac{1}{128\pi} \frac{\luv^2}{\l4^2}  .
\end{equation}

On the other hand, the number of light 
species in a typical Grand Unified Theory is roughly 
the number of gauge bosons. It then follows from    
Eq. (\ref{defb}) that $b \gsim n /(100\pi^2)$ and substituting 
this relation into Eq. (\ref{bcutoff}) leads to the condition
\begin{equation}
\label{saturate?}
\luv \gsim \sqrt{n} \l4   .
\end{equation}
This bound for the cutoff is in agreement with 
independent perturbative \cite{pertcut} and non-perturbative 
\cite{nonpertcut,dvaliinfo} analyses.

Eq. (\ref{triality}) also provides a way of quantifying the model-dependent 
GUP parameter (and consequently the four-dimensional UV 
cutoff) directly in terms of five-dimensional 
length scales. This is interesting from the holographic perspective,  
since the strong coupling scale in four dimensions 
is determined by the characteristic scale of the 
higher-dimensional gravity \cite{susswitten,dvaliinfo}. 
For $\AdS5$, one would therefore expect that 
Eq. (\ref{triality}) should be consistent with 
the condition $\luv  \simeq \ell$. 
To verify this, we substitute
$\luv \simeq 2 |\alpha| \l4$ into Eq. (\ref{triality}) to deduce that 
\begin{equation}
\label{expect}
\frac{\luv^2}{\l4^2} = \frac{\ell^3}{\planck5^3}
\end{equation}
The condition $\luv \simeq \ell$ then simplifies this expression to  
a relation between the four- and five-dimensional Planck scales, 
$\ell \simeq \planck5^3 /\l4^2$. This is precisely the dependence that  
arises from a direct Kaluza-Klein reduction from five to four 
dimensions for a horospherical brane embedded in $\AdS5$ 
\cite{RSmodel,gubser1}. 

The scale $\ell$ also represents the scale below which the $\AdS5$ curvature 
becomes negligible and gravity becomes effectively
five-dimensional. In this context, 
the correspondence (\ref{triality}) 
implies that the minimum measureable length can be identified with the
scale at which four-dimensional physics breaks down.   
A similar conclusion was arrived at in a different 
context by a direct investigation of quantum systems 
with one extra dimension compactified on a circle \cite{mwy}.  

Before concluding, we should discuss the validity of the 
assumptions we have made. Firstly, the GUP we have invoked 
in Eq. (\ref{GUP}) is heuristic, in the sense that it has not been derived 
from first principles. Thus, although it ultimately 
leads to the trace-anomaly Friedmann equation, it is probable that 
the thermodynamic approach we have developed does not incorporate all the 
corrections that are expected to arise to the gravitational action. 
In particular, one would expect graviton loops to generate 
higher-order corrections that also lead to further  
$H^4$-correction terms in the Friedmann equation. Thus, 
our derivation is necessarily incomplete. 

Another key assumption we made was that the apparent horizon should 
vary sufficiently slowly with respect to cosmic time, 
such that its area remains effectively constant during the 
time interval it takes for a quantum particle to be emitted. 
This is equivalent to assuming 
a quasi-de Sitter (inflationary) expansion and can be realised 
by a slowly-rolling, self-interacting scalar field (as we implicitly assumed). 
Such a field can be
interpreted as the perfect fluid responsible for the energy-momentum flux 
through the horizon, Eq. (\ref{energyflow}). 
(A self-interacting scalar field minimally coupled to Einstein gravity 
is dynamically equivalent to a perfect fluid in a spatially isotropic 
universe.) On the other hand, 
{\em a posteriori} such an assumption is not necessary, since
the thermodynamic Friedmann equation (\ref{thermoFriedmann}) admits 
solutions of the form
\begin{equation}
\label{quadratic}
H^2 = \frac{1}{\alpha^2l_4^2} \left[ 1 + \epsilon 
\sqrt{1-\frac{16\pi l_4^4 \alpha^2}{3} \rho} \right] \, ,
\end{equation}
where $\epsilon =\pm 1$. The $\epsilon =+1$ root is essentially 
the Starobinsky \cite{starobinsky}
model of inflation driven by an $R^2$-correction to the 
Einstein-Hilbert action, where $R$ is the Ricci curvature scalar
\cite{gub}. In this 
case, one may regard the fluid crossing the apparent horizon as 
the conformal (massless) matter fields. 

In conclusion, we have considered the effect of the generalized
uncertainty principle on the entropy of the apparent horizon and found that 
the trace-anomaly Friedmann equation 
can be derived from the first law of thermodynamics
when Eq. (\ref{triality}) is satisfied. For a generic CFT, such 
a correspondence implies that perturbative theory should break down 
at a scale $\luv \gsim  \sqrt{n} \l4$
in a theory containing ${\cal{O}} (n)$ light species.
For a CFT with a holographic dual,
it identifies the minimum
measureable length in four dimensions as the 
scale where physics becomes effectively five-dimensional. It is also 
worth remarking that trace-anomaly inflation provides an excellent 
fit to the Planck satellite observations of the cosmic microwave background 
anisotropies \cite{planck}.

\end{document}